# On-chip unidirectional waveguiding for surface acoustic waves along a defect line in a triangular lattice

Yun Zhou[1,4], Naiqing Zhang[1,4], Dia'aaldin J. Bisharat[2], Robert J. Davis[2], Zichen Zhang[3], James Friend[1], Prabhakar R. Bandaru[1, 2, 3] and Daniel F. Sievenpiper[2, 3]

1 Department of Mechanical Engineering, University of California, San Diego, La Jolla 92093-0411 CA, USA

2 Department of Electrical Engineering, University of California, San Diego, La Jolla, 92093-0411 CA, USA

3 Program in Materials Science, University of California, San Diego, La Jolla 92093-0411 CA, USA

4 These authors contribute equally to the paper.

Email: jfriend@eng.ucsd.edu, pbandaru@eng.ucsd.edu and dsievenpiper@eng.ucsd.edu

The latest advances in topological physics have yielded a rich toolset to design highly robust wave transfer systems, for overcoming issues like beam steering and lateral diffraction in surface acoustic waves (SAWs). However, presently used designs for topologically protected SAWs have been largely limited to spin or valley-polarized phases, which rely on non-zero Berry curvature effects. Here we propose and experimentally demonstrate a highly robust SAW waveguide on lithium niobate (LiNbO$_3$), based on a line defect within a true triangular phononic lattice, which instead employs an intrinsic chirality of phase vortices and maintains a zero Berry curvature. The guided SAW mode spans a wide bandwidth ($\frac{\Delta\omega}{\omega_{\text{center}}}$ =9.17%) and shows confinement in the lateral direction with 3 dB attenuation within half of the unit-cell length. SAW routing around sharp bends has been demonstrated in such waveguide, with less than ~4% reflection per bend. The waveguide has also been found robust for defect lines with different configurations. The fully on-chip system permits unidirectional SAW modes that are tightly bound to the waveguide, which provides a compact footprint ideal for miniaturization of practical applications and offers insight into the possibility of manipulating highly focused SAW propagation.

## Introduction

Surface acoustic waves (SAWs) have been appealing for applications requiring precise on-chip manipulation[1–3] of particles and fluids. In acoustofluidics[4], for example, SAWs are widely used for life science-related applications such as cell mixing[5,6], cell sorting[7,8], and tissue engineering[9,10]. However, beam steering[11] due to anisotropy of the piezoelectric substrate, and lateral diffraction of SAWs cause energy dissipation, resulting in degraded performance. SAW waveguides[12] can be utilized for their capability to confine and control wave propagation to enhance the performance of these devices. SAWs have also been proposed as an interface for probing and controlling elementary excitations in condensed matter[13–18]. In large-scale, high-fidelity on-chip phononic quantum networks, unidirectional and low-loss SAW waveguiding is desired. SAW filters are also key components for wireless devices[19,20], given their small form factors, inexpensive manufacturing, and high Q factor. As modern multi-band systems continue to shrink in size, it is becoming increasingly important to miniaturize such SAW filters without sacrificing performance. Having a design scheme to steer acoustic wave as desired while suppressing backscattering and wave routing capability would provide the possibility of further reducing the size of SAW RF filters.

A promising method to create unidirectional and backscatter immune waveguides are designs inspired by the field of topological insulators (TIs) broadly based upon the quantum Hall effect[21,22]. In fermionic settings, such a phase is realized by breaking time-reversal symmetry (TRS), and has been extended to bosons by introducing external rotational forces for many photonic[23–28] and mechanical/acoustic systems[29–36]. However, the realization of topologically protected modes in the technologically relevant case of SAW has thus far proved elusive. Intrinsic TRS breaking of ferromagnetic materials can also lead to non-reciprocal SAW propagation[37–39], due to the different absorption in the $+k$ and $-k$ directions, but the difference is usually small. Moreover, broken TRS requires magnetism, making these devices undesirable in real integrated circuits. Another way to achieve acoustic non-reciprocity is to use nonlinear effects[40–43]. However, the manipulation of propagation direction remains challenging in these nonlinear systems.

On the other hand, TIs based on lattice symmetry breaking are reciprocal and protected by TRS, which are passive and have been proven to guide robust and unidirectional waves in many bosonic systems. For example, tuning the inter- and intra- cell coupling while maintaining $C_{6v}$ symmetry in honeycomb lattices[44–46], or breaking the *z* directional mirror symmetry in bianisotropic materials[47–49], can introduce pseudospins that mimic the quantum spin Hall effect[50–53]. Likewise, unidirectional valley TI waveguides can be constructed by breaking inversion symmetry in a honeycomb lattice[54]. These discoveries lead to extending the idea of reciprocal TIs to on-chip phononic devices[55–57]. However, due to the lattice symmetry requirement, most of the existing on-chip designs are suspended structures for bulk acoustic wave or Lamb waves, where an easy implementation for SAW is still lacking.

Here, we report a scalable, non-suspended, fully integrated, reciprocal, and unidirectional on-chip SAW waveguide upon lithium niobate ($LiNbO_3$). In our implementation, the unidirectional SAW waveguide is created by a defect boundary in a triangular phononic lattice. We demonstrate that despite the triangular lattice having a vanishing Berry curvature, the intrinsic phase vortices embedded in the triangular lattice give rise to unidirectional wave transport. We prove that, compared to valley TIs, our simpler waveguide structure shows a much better lateral confinement without sacrificing the directionality, which is beneficial for integration into SAW devices in on-chip applications. The confined, robust, and unidirectional SAW routing phenomenon has been verified in experiment, which demonstrates that the SAW overcomes the limitation of beam steering in a piezoelectric substrate and is capable of making sharp turns along the defect line waveguide with low reflection and loss. By incorporating the proposed waveguides, there is now a possibility of wave propagation in any chosen direction, which can benefit many SAW applications.

## Results

**Unidirectional SAW waveguide in a triangular lattice with zero Berry curvature**

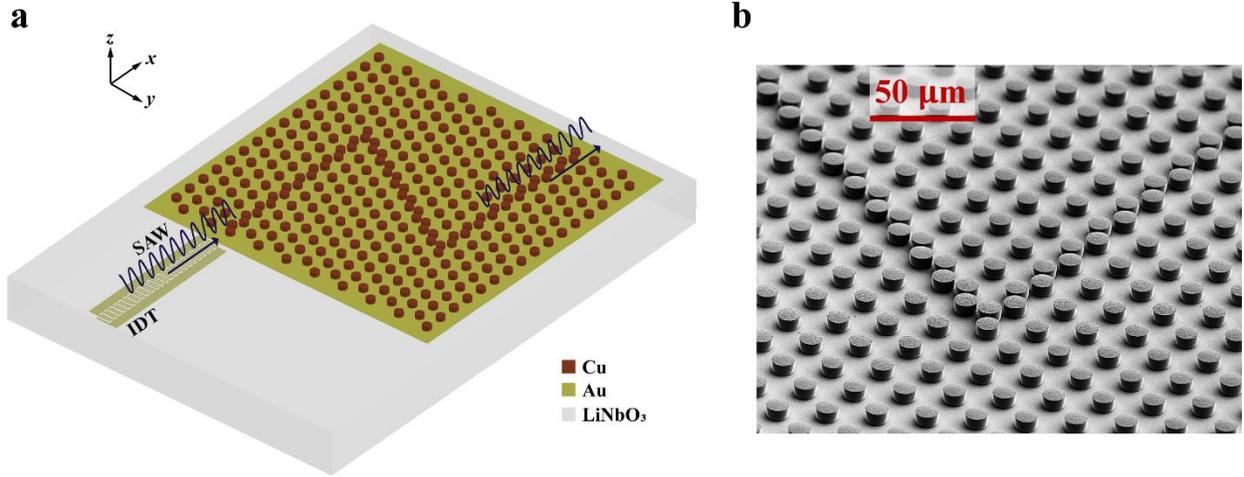

**Figure 1 a** Schematic of the proposed defect-line SAW waveguide composed of phononic crystal of Cu pillars arranged in a triangular lattice on a 127.68° Y-rotated X-propagating LiNbO$_3$ wafer. The incident SAW in $x$ direction is provided by a broadband chirped IDT. **b** SEM image with a zoomed-in view of the fabricated SAW waveguide. The Cu pillars are 11.5 μm in diameter and 6.2 μm in height, with lattice constant $a$=24 μm, grew on a 400 nm Au seed layer on top of a 500 μm LiNbO$_3$ wafer.

Our proposed SAW waveguide is formed by a defect line in a triangular array of copper pillars on 127.68° Y-rotated X-propagating LiNbO$_3$ wafer, as shown in **Fig. 1a**. The entrance and exit ports of the waveguide are aligned with the X direction of the LiNbO$_3$ wafer. The lattice made of identical copper pillars introduces periodic modulation for SAWs, which produces in SAW dispersion bands and bandgaps. **Fig. 1b** shows a scanning electron microscopy (SEM) image of the fabricated SAW waveguide, where the copper pillars were grown onto the LiNbO$_3$ substrate through electrochemical deposition. To study the SAW propagation, a broadband interdigital transducer (IDT) with a narrow aperture (see **Methods**) is fabricated on the same wafer and excites SAWs in the X direction from the entrance port.

It has previously[58] been observed that a Dirac degeneracy at the $K$ point for SAWs occurs in a honeycomb lattice consisting of metallic pillars on LiNbO$_3$ (also shown in **Fig. 2a** and **d**). When differentiating one pillar from the other in the unit cell by shrinking the diameter of pillar B, as shown in **Fig. 2b,** the structure becomes a valley TI[59]. The $C_{6v}$ symmetry of the honeycomb lattice is reduced to $C_{3v}$, which lifts the Dirac degeneracy at the $K$ point in the honeycomb lattice and forms a SAW bandgap, as shown in **Fig. 2e**, where topologically protected valley edge states for SAWs are expected to be found[59]. Further shrinking the pillar B diameter to zero, the number of copper pillars in one unit cell reduces from two to one, and the honeycomb lattice is transformed into a triangular lattice[60], as shown in **Fig. 2c**. Since the SAW modes are supported by the mechanical resonances of the copper pillars (see **Supplementary Information** section 4), reducing the number of pillars by half reduces the number of SAW modes under the sound cone by half. As shown in **Fig. 2f**, a larger SAW bandgap then forms from 73.08 MHz to 88.13 MHz.

We observed intrinsic phase rotation in the triangular lattice, while, in contrast to the valley TI, its Berry curvature vanishes. The phase distribution map of the out-of-plane displacement field $u_z$ of the highlighted bands at the high symmetry point $K$ of the Brillouin zone for the honeycomb lattice, valley SAW TI, and triangular lattice are shown in **Fig. 2g, h and i**, respectively. It can be observed that the phase shows greater uniformity close to the pillars. In the case of the valley SAW TI, as shown in **Fig. 2h**, the relative position of the pillars leads to two vortices with opposing directions from a $2\pi$ phase rotation in the lattice: one showing a counterclockwise vortex at the center of three pillars arranged in upwards triangles, and one showing a clockwise vortex at the center of three pillars arranged in downward triangles. These vortices indicate the intrinsic presence of circular-polarized orbital angular momentum (OAM) and a chiral property for $u_z$ throughout the bulk of the valley TI lattice[61,62]. The OAM waves with opposite signs suggest unidirectional interfacial modes are supported when the directionality is enforced rather than cancelled at a boundary or interface. The phase plot in **Fig. 2i** shows that the triangular lattice with only one pillar still maintains the intrinsic OAM waves that occur in a nontrivial valley TI, implying unidirectional confined edge modes would still be supported in the bandgap in such a triangular lattice. However, despite the similarity, the triangular lattice maintains a $C_{6v}$ point group symmetry, while the lattice symmetry in a valley structure is reduced to $C_{3v}$. For the valley TI, inversion symmetry results in $\pm\pi$ Berry phase accumulation around $K$ or $K'$, as shown in **Fig. 2j**. By contrast, with no TRS breaking or inversion symmetry breaking, the Berry curvature vanishes everywhere in the Brillouin zone for the triangular lattice (See **Supplementary Information** section 1). This is illustrated in **Fig. 2k**, where the Berry curvature is zero[63] around $K$, a clear contrast to the case of the valley structure.

The spatial arrangement of the phase vortices suggests a gauge dependance on the existence of edge modes along the border. This can be likened to the topological crystalline insulator phases found in Kagome crystals[64–66], where the underlying symmetry is determined by the choice of unit cell. We can imagine the limit of a "breathing" Kagome unit cell being equivalent to a triangular lattice, with each lattice cite partially overlapping with the neighboring cells[66,67]. Unlike Kagome crystals, however, the physical realization given here maintains the $C_{6v}$ rotational symmetry for suitable choices in unit cell. Despite this, it can be seen (see the **Supplemental Information** section 2 and 3) that for the unit cell definitions that result in $C_{3v}$ edges along finite boundaries (as are studied here) the edge modes can be well described by a nontrivial symmetry indicator[68] that describes the effect of the phase vortex. The existence of unidirectional modes is therefore a direct consequence of the real space behavior of the finite crystal along certain boundaries, rather than the reciprocal space influence of valley-based effects. The symmetry indicator provides a direct measure of these real space effects, which for our system is visually represented in the phase vortices seen in **Fig. 2i**.

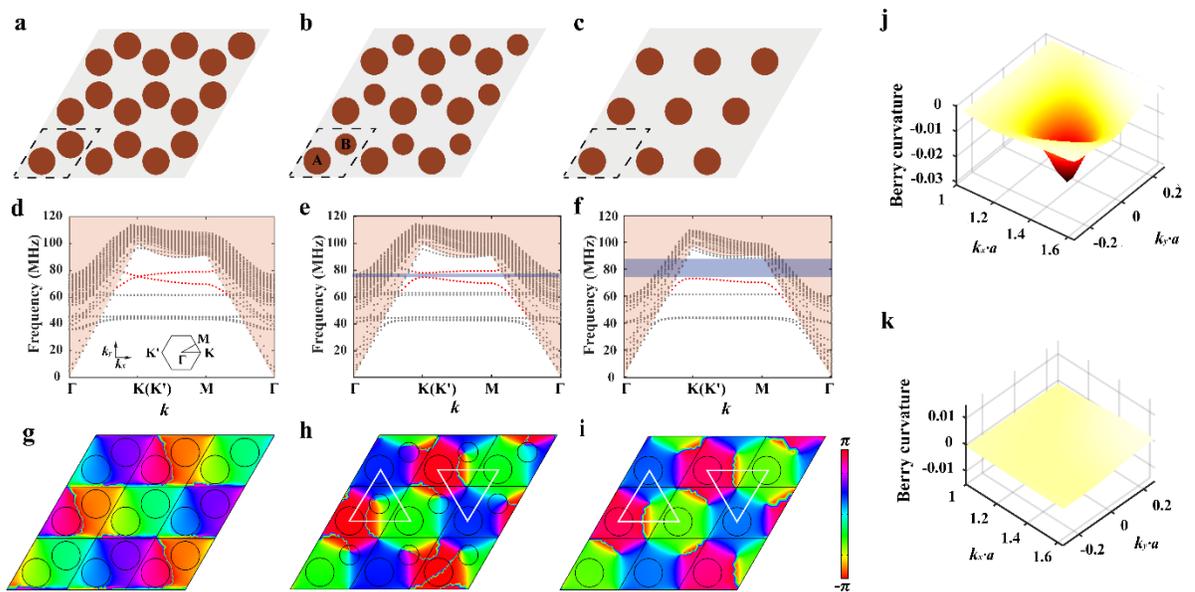

**Figure 2 a** Honeycomb lattice, **b** topological valley TI with $r_B = 0.8 r_A$ and **c** a triangular lattice of Cu pillars of 11.5 μm diameter and 6.2 μm height on 500 μm LiNbO$_3$ substrate. Calculated band dispersions along Γ-K (K′)-M-Γ direction for **d** the honeycomb lattice, **e** the valley TI, and **f** the triangular lattice. Bandgaps for SAWs due to symmetry breaking are shaded in blue. Bulk acoustic wave bands are shaded in orange. We focus on bands highlighted in red dashed lines this paper. Simulated phase maps for the **g** honeycomb lattice, **h** valley TI and **i** the triangular lattice at K. Phase plots for the valley TI and the triangular lattice shows two vortices: one at the center of the downward triangles, and one at the center of the upward triangles. **j** Berry curvature for the triangular lattice around K $(\frac{4}{3}\frac{\pi}{a}, 0)$. The Berry curvature is zero thoroughout the BZ with no accumulation around K, which shows a clear contrast to **k** that of the valley TI.

## Unidirectional SAW edge states

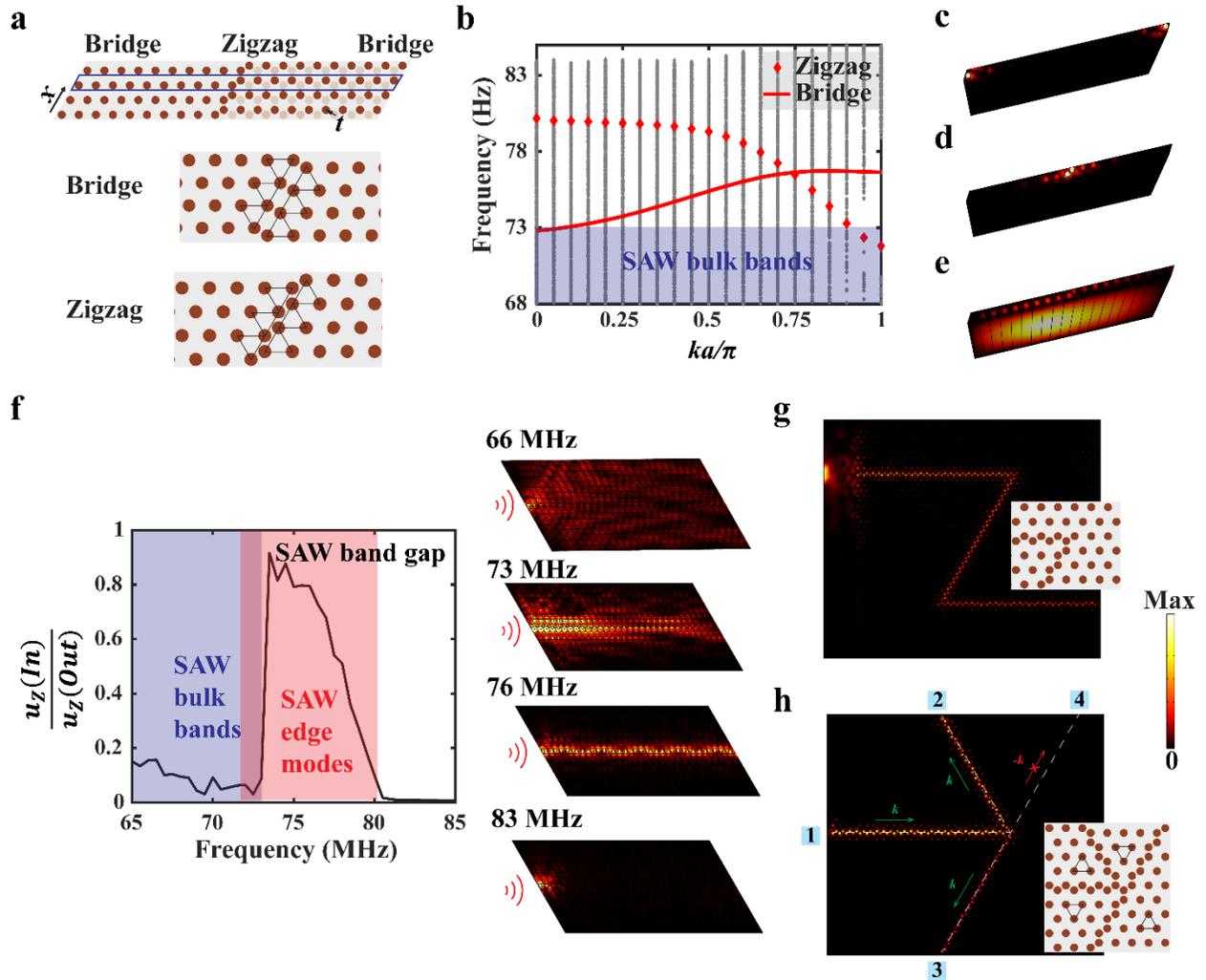

**Figure 3 a (top)** A defect-line SAW waveguide created by shifting one half domain of the triangular phononic crystal of $\vec{t} = \frac{\sqrt{3}}{3}a$ in the direction perpendicular to the waveguide. The light red color shows the positions of the pillars before the shift. A ribbon super cell is highlighted in blue, with a zigzag interface in the middle and bridge interface at the edges (as periodic boundary condition is applied in *x* direction. **(bottom)** Bridge and zigzag defect lines in triangular lattice. The upward triangles Δ and downward triangles ∇ indicate opposite phase vortices. **b** Band dispersions for the ribbon super cell. **c** Eigen-displacement $u_z$ for the edge mode confined at the bridge interface, corresponding to the edge mode in solid red line in **b**. **d** Eigen-displacement $u_z$ for the edge mode confined at the zigzag interface, corresponding to the dashed red line in **b**. **e** Eigen-displacement $u_z$ for the bulk acoustic modes in the SAW bandgap, corresponding to the grey dots in **b**. **f (left)** $u_z$ amplitude at the exit port of a straight defect-line waveguide of 64*a* length normalized by $u_z$ amplitude at the entrance port from 65 MHz to 85 MHz. **(right)** $u_z$ field plots for the straight defect-line waveguide at 66 MHz (within SAW bulk bands), 73 MHz (edge mode within SAW bulk bands), 76 MHz (edge mode within SAW bandgap) and 83 MHz (within SAW bandgap). **g** Simulated $u_z$ field for a Z-shaped waveguide with an excitation of 76 MHz. **h** A "magic T" junction for the defect-line SAW waveguide. The "magic T" divide the domain into 4 parts with 60°, 60°, 60°and 180° angles at the junction. The four sub-domains are denoted by the Δ and ∇ respectively. The excitation is at 76 MHz at port 1.

We prove through numerical simulations the existence of unidirectional SAW edge states, despite zero Berry curvature. We construct a defect-line waveguide in the lattice where the phase on its two sides shows opposite vortices that enhances the direction of the energy flow at the defect line (see **bottom** figure in **Fig. 3a**). Here, a waveguide is created by shifting the right half domain of the triangular lattice $t = \frac{\sqrt{3}}{3}a$ in the direction perpendicular to the zigzag defect line, as illustrated in the top figure of **Fig. 3a**. It can be observed that the domain on the left of the waveguide is truncated such that the pillars are arranged in downward triangles along the boundary, while along the right side of the waveguide the pillars are arranged in upward triangles. The phase vortices are opposite on the two sides, which leads to energy flux constructively adding up in one direction, giving rise to unidirectional SAW transport at the defect line. To find guided modes along the interface, we calculate the band dispersion of a ribbon supercell with a zigzag interface in the middle and bridge interface on the edges, as highlighted in the top figure of **Fig. 3a**. Periodic boundary conditions are applied in both the *x* direction and the direction along the waveguide, and the calculated band diagram is plotted in **Fig. 3b**. The two bands highlighted in red fall mostly into the SAW bandgap of the triangular lattice, with their corresponding eigen $u_z$ amplitude fields shown in **Fig. 3c** and **Fig. 3d**, indicating the existence of the edge states at both the zigzag and the bridge interfaces. The edge mode associated to the zigzag interface spans a wider frequency range (71.8 MHz to 80.18 MHz) compared to the edge mode confined at the bridge interface (72.8 MHz to 76.65 MHz). There are modes for bulk acoustic waves that also fall in the SAW bandgap, as shown by the grey dots in the band diagram plot, which are above the sound cone in **Fig. 2b** and are obtained here by zone folding when considering the supercell instead of the primitive unit cell. **Fig. 3e** shows the $u_z$ of one of the modes in grey dots, which clearly shows bulk wave behavior. Since these modes propagate into the bulk at a higher velocity, we expect them to have minor coupling with our SAW edge modes. **Fig. 3f** shows the $u_z$ amplitude at the exit port of a straight zigzag defect-line waveguide of 64*a* length normalized by the $u_z$ at the entrance port, excited by a point source at the entrance. It can be observed that the SAW waveguide has high transmission from 73.08 MHz to 80.18 MHz, where the edge mode happens. For frequencies below 73.08 MHz (within the SAW bulk bands), SAWs radiate throughout the whole surface, while for frequencies above 80.18 MHz the SAW bandgap prohibits SAW propagation, leading to low transmission for those frequency ranges. **Fig. 3g** shows a driven mode simulation for a zigzag-type of defect-line SAW waveguide with two 120-degree sharp turns, demonstrating robust SAW waveguiding with little reflection. To verify the unidirectionality, we constructed "magic T" junctions as shown in **Fig. 3h**. These "magic T" junctions consist of four defect-line SAW waveguides that separate the domain into 4 parts. When sending a wave into port 1, the excited SAW will propagate in the direction in which the wave sees the upward triangles on its left and the downward triangles on its right. As shown in **Fig. 3h**, the SAW excited at port 1 couples to port 2 and 3 but not port 4, as the waveguide connected to port 4 only supports SAW in the opposite propagation direction. As such, the edge states at the defect-line waveguide are unidirectional.

## Confinement and robustness of the defect-line SAW waveguide

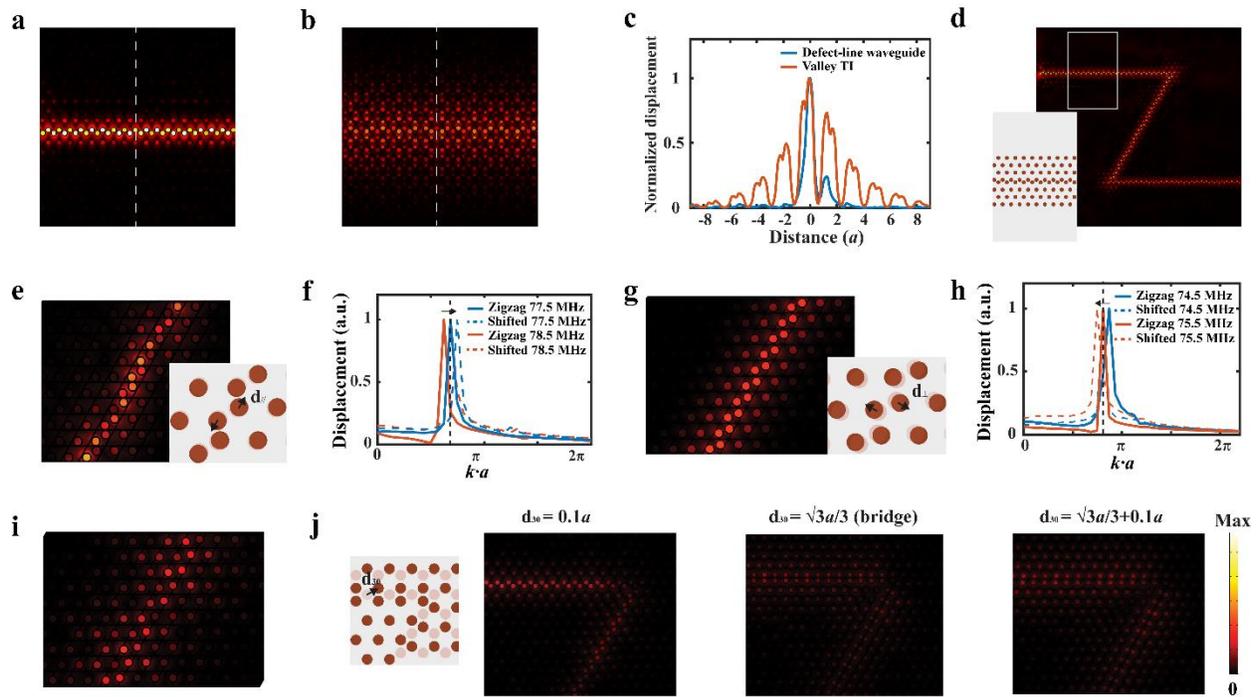

Figure 4 **Confinement and robustness of the defect-line waveguide.** **a** The proposed defect-line waveguide excited at 76 MHz. **b** the valley TI with $r_B = 0.8r_A$ excited at 76 MHz. **c** Normalized displacement distribution in the direction perpendicular to SAW propagation for the defect-line waveguide and the valley TI. **d** Z-shape defect-line waveguide with only 3 rows of pillars on both sides of the waveguide. The inset shows a zoomed in view of the waveguide. **e** a zigzag waveguide with a shift $d_\parallel = 0.0425a$ along the waveguide excited at 76 MHz. **f** Spatial FFT for $u_z$ along the shifted waveguide in **e** compared to that of a zigzag waveguide at excited 77.5 MHz and 78.5 MHz. The SAW travels at 77.5 Hz along the zigzag waveguide shares the same wavevector as the SAW travels along the shifted waveguide at 78.5 MHz, indicating that the edge modes move to higher frequencies. **g** A zigzag waveguide with a shift $d_\perp = 0.05a$ perpendicular to the waveguide excited at 76 MHz. **h** Spatial FFT for $u_z$ along the shifted waveguide in **g** compared to that of a zigzag waveguide at 74.5 MHz and 75.5 MHz. The SAW travels at 75.5 Hz shares the same wavevector as the SAW travels along the shifted waveguide at 74.5 MHz, indicating that the edge modes move to lower frequencies towards the bulk SAW bands. **I** A zigzag waveguide with a shift $d_\perp = 0.15a$ perpendicular to the waveguide excited at 74.5 MHz. **k** Bent zigzag

waveguides with a shift $d_{30} = 0.1a, \frac{\sqrt{3}}{3}a$ and $(\frac{\sqrt{3}}{3} + 0.1)a$ 30° to the waveguide excited at 76MHz, from left to right, respectively.

We compare the confinement of the proposed defect-line waveguide with a valley TI. The displacement fields at 76 MHz are shown in **Fig. 4 a** and **b** for our defect waveguide in contrast to a valley TI (**Fig. 2 b**), respectively, with the displacement perpendicular to the waveguide plotted in **Fig. 4 c**. The defect-line waveguide shows a much faster decay along the orthogonal direction, with 3 dB decay in $0.264a$, and 20 dB decay in $0.533a$. The displacement for the valley TI is more spread out, with 3 dB decay in $1.324a$, and 20 dB decay in $3.140a$. The capability of confining SAW in a narrow region allows us to construct a defect-line waveguide with fewer unit cells in the orthogonal direction. **Fig. 4 d** demonstrates that a Z-shape defect-line waveguide containing only three unit cells on either side of the interface still functions as expected.

The waveguides we discussed above are constructed by shifting the pillars to form a perfect zigzag or bridge grain boundary defect line in a triangular lattice. We explored further the configuration of the defect line and how it affects the confinement of the SAW edge states. **Fig.4e** shows a waveguide with the left and right domains of the zigzag interface both shifted in the direction parallel to the interface by $d_\parallel = 0.0425a$, while **Fig.4g** illustrates a waveguide with the left and right domains of the zigzag interface both shifted in the perpendicular direction away from the interface by $d_\perp = 0.05a$. It can be observed that in both cases the waveguides support SAW propagation, showing the robustness of the edge modes. However, from the spatial FFT for the two cases, as shown in **Fig.4f** and **Fig. 4h**, respectively, the edge mode is shifted to higher or lower frequencies. **Fig. 4i** shows the case with perpendicular shifting of $d_\perp = 0.15a$, where the SAW is still guided through the interface, but is less confined to the waveguide. This is because shifting the two domains away from each other will reduce the coupling of the two opposite phase vortices at the interface and push the edge modes more towards the SAW bulk bands, leading to a less confined interfacial mode. **Fig. 4j** shows 120-degree bent defect-line waveguides with half of the domain shifted $d_{30}$ in the direction 30° to the zigzag interface. The field plots in **Fig. 4j** are for the cases when $d_{30} = 0.1a, \frac{\sqrt{3}}{3}a$ and $(\frac{\sqrt{3}}{3} + 0.1)a$, respectively. Similarly, as the shifting distance $d_{30}$ increases, the SAW also becomes less confined. Note that when $d_{30} = \frac{\sqrt{3}}{3}a$, the waveguide becomes a perfect bridge interface. As suggested by Fig. **3a**, the edge mode of a bridge interface is closer to the SAW bulk band and with slower propagation velocity compared to that of a zigzag interface, resulting in worse confinement.

**Experimental observation of the SAW waveguiding**

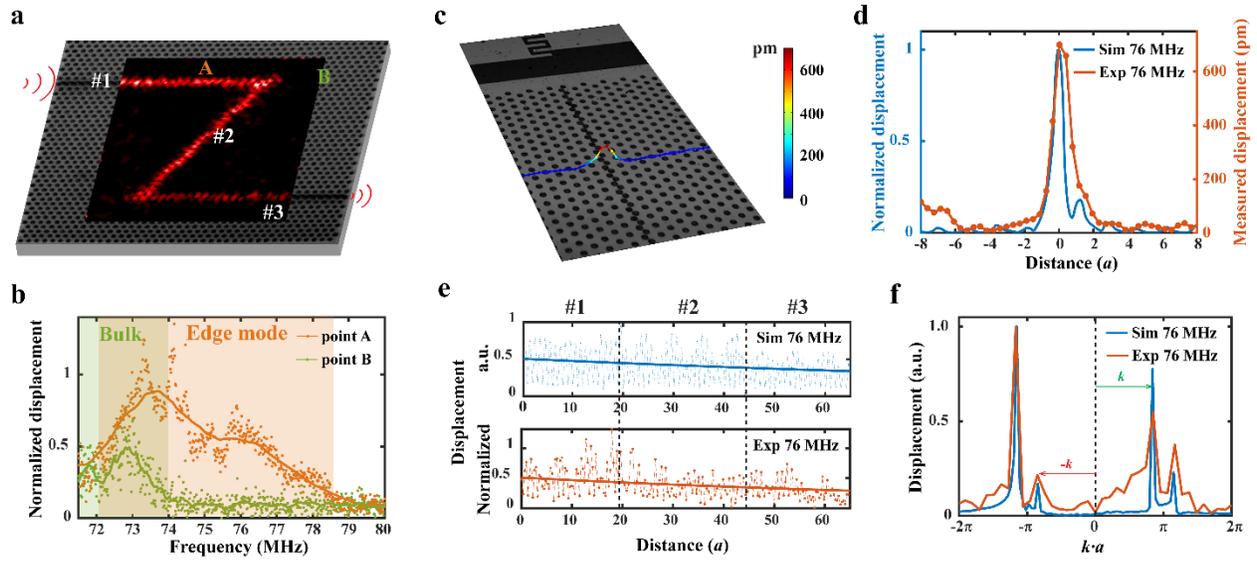

**Figure 5 a** Measured $u_z$ field for a Z-shaped SAW defect waveguide. The Z shaped waveguide consists of three segments with $26a$ length for each segment, and two 120º sharp turns. The SAW was excited by a broadband IDT with bandwidth from 35 MHz to 90 MHz. The $u_z$ is imaged by the LDV over a 752.15 μm by 612.80 μm rectangular region. The $u_z$ field shown is at 76 MHz. **b** Comparison of $u_z$ for point A on the waveguide and point B away (illustrated in **a**) from the waveguide over the frequency range from 71.5 to 80 MHz, respectively. Measured data is plotted in dots, with the moving average shown in solid line. **c** Confinement measurement at 76 MHz. **d** $u_z$ along the direction perpendicular to the propagation direction at 76 MHz for the simulation and the experiment, respectively. **e (top)** Simulated $u_z$ along the waveguide at 76 MHz (dashed line) and fitted exponential decay (solid line). (**bottom**) Measured $u_z$ along the three segments of the waveguide (illustrated in **a**) at 76 MHz (dashed line) and fitted exponential decay (solid line). The $u_z$ in **b** and **e** are normalized by the distance between the IDT and the entrance port of the waveguide. **f** Spatial FFT at 76 MHz for the simulation and the experiment.

Tightly confined SAW guiding along the proposed waveguide was experimentally demonstrated in our device, as shown in **Fig.5**. The out-of-plane displacement field $u_z$ is measured by a laser Doppler vibrometer (LDV, UHF-120, Polytec). To eliminate possible spurious mode interference from bulk acoustic waves, the back side of the LiNbO$_3$ was roughened. We designed a chirped IDT with a wide bandwidth from 35 MHz to 90 MHz (see **Methods**) to excite the SAW and observed edge states propagating through a Z-shaped interface with two 120-degree sharp turns. **Fig. 5a** depicts the measured out-of-plane SAW displacement field $u_z$ at 76 MHz where a clear SAW confinement and transport along the defect-line waveguide is shown. To determine the SAW edge mode bandwidth, we compared two points on the device: point **A** on the waveguide and point **B** in the bulk of the triangular crystal, as shown in **Fig. 5a**, and compared their $u_z$ vs. frequency, as shown in **Fig. 5b**. Here, the $u_z$ is normalized by the displacement directly in front of the source IDT. It can be clearly seen that away from the waveguide (point **B**) the $u_z$ goes to nearly zero after 74 MHz, which indicates a SAW bandgap for the triangular lattice above 74 MHz. On the other hand, the displacement profile of point A on the waveguide shows a clear bandwidth up to 78.5 MHz, which proves that our edge mode exists in the bulk band of the SAW from 74 MHz to 78.5 MHz. We have also observed guided SAW below the bulk band from 72 MHz to 74 MHz (see **Supplementary Information** section 6), which is also confirmed by **Fig. 3a**. However, these modes coexist with bulk SAW modes in the background and are less confined. At higher frequencies we note a reduced bandwidth than in simulation, owing to the band edges resulting in flat dispersion, thereby increasing their attenuation, and complicating direct observation. To show the confinement of the edge mode in the SAW bandgap, we measured the $u_z$ in the direction perpendicular to the waveguide, as shown in **Fig. 5c**. **Fig. 5d** shows that the measured $u_z$ agrees with simulation at 76 MHz, and a 3dB decay of the displacement amplitude within $0.509a$ is observed, implying that the mode is highly confined to the interface.

A decay in the displacement amplitude along the waveguide was observed in both the simulation and the measurement (see **Supplementary Information** section 5). The measured $u_z$ amplitude for the three segments of the Z-shaped waveguide at 76 MHz is shown in the bottom figure of **Fig. 5 e**, with the simulation result shown in the top figure of **Fig.5 e**. If we assume negligible reflection at each sharp turn and fit the decay trend of the $u_z$ with an exponential decay $Ae^{-\alpha d}$, where $A$ is the displacement amplitude, $\alpha$ is the decay coefficient, and $d$ is the distance SAW travels along the waveguide, we find $\alpha_{sim}$ =0.00620/$a$, and $\alpha_{exp}$ =0.00862/$a$, with the fitted exponential curves plotted in solid lines in **Fig. 5 e**. This indicates that 3dB loss happens at a distance of $55.9a$ from the entrance port in the simulation, and $40.2a$ in the measurement.

We quantitively studied the reflection of the SAW and demonstrated there is indeed little reflection at the 120-degree sharp turns. We took the spatial FFT of the $u_z$ along the first segment of the waveguide before the first sharp turn and looked at the wavevectors, as shown in **Fig. 5f**. The wavenumber components in **Fig. 5f** show a finite value for the negative wavenumber $-k$ within the first BZ ($-\pi$ to $\pi$) for both the simulation and the experimental results, indicating there is a small reflection at the 120-degree bends. A higher order component for the same wavevector outside of the first BZ is also observed. We took the average ratio of the displacement component for the $k$ and the $-k$: $r = \frac{u_{z,-k}}{u_{z,k}}$ in the 1$^{st}$ and 2$^{nd}$ BZ as the reflection coefficient of the SAW and obtained $r_{sim} = 0.224$ for the simulation, and $r_{exp} = 0.385$ for the measurement. We consider the 3 segments of the Z-shaped waveguide of same acoustic impedance R, so that the SAW energy flux can be expressed as $|u_z|^2/R$. Assuming all the SAW are either reflected or transmitted at the bend, the transmission for the Z-shaped waveguide can be estimated as $t = \sqrt{1-r^2}$, which is $t_{sim}$= 0.975 for the simulation, and $t_{exp}$= 0.923, for the measurement. The experimental result indicates that less than 8% of the SAW is reflected in the direction opposite to the incident direction by the two 120-degree bends, proving the directionality of the waveguide. The discrepancies between the simulation and measurement are due to inevitable damping in the sample (e.g., the electroplated copper pillars) and fabrication errors, which are difficult to be precisely simulated.

In summary, we have developed a fully integrated on-chip topological SAW unidirectional waveguide, based on defect lines in a triangular phononic lattice constituted from metallic pillars. Different from spin or valley topological structures, the phononic lattice is trivial with regard to the Berry curvature and is instead maintained by the phase vortex distribution in real space. With half of the total numbers of lattice points needed compared to a valley TI, our proposed SAW defect-line waveguide shows better confinement in the lateral direction. The confined SAW reduces the number of unit cells needed to construct the waveguide, making it possible to fit multiple of such SAW waveguides on a small chip. Our experiments, supported by simulation results, show successful SAW confinement and routing with small reflection around sharp bends in the propagation path. Combined with an optimal cut for arbitrary propagation of SAW on a LiNbO$_3$ substrate[69], we anticipate the ability to pass SAWs along any direction. Different configurations at the defect boundary have been studied, showing that our waveguide is robust for different configurations, with the zigzag interface having the widest bandwidth and greatest confinement. Altering the defect boundary shifts the propagation frequency up or down, which can be potentially used to manipulate SAW propagation based on a small change in frequency. These results demonstrate the value of this system for further scientific investigations and device development, such as precise control of removing cells locally from culture surface[70], multistage cell sorting, high pressure SAW pumping[71] and acoustic streaming[1].

## Methods

### Sample preparation

We fabricated chirped interdigital transducers (IDTs) on 500 μm thick, double-side polished 128° Y-rotated x cut lithium niobate (LN, Precision Micro-Optics Inc., Burlington, MA, USA) for surface acoustic wave generation and propagation. Finger widths and finger gaps varying from 26 μm to 11 μm were selected for an operating frequency of 40-90 MHz (from $f = v/\lambda$) to define each IDT, comprised of twenty-five simple finger pairs with finger and linearly distributed gap widths. Standard UV photolithography (using AZ 1512 photoresist and AZ 300MIF developer, MicroChem, Westborough, MA, USA) was used alongside sputter deposition (Denton 18, Denton Vacuum, NJ, USA) and lift-off processes to fabricate the 10 nm Cr / 400 nm Au IDTs and seed layer upon the LN substrate[72,73]. The second layer structure with a thickness of ~15 μm for pillar growth was fabricated via standard UV laser-written photolithography with alignment to the first layer of the IDT structure (using AZ 12XT-20PL-10 photoresist and AZ 300MIF developer, MicroChem, Westborough, MA, USA) (MLA 150, Heidelberg Instruments, Heidelberg, Germany). A dicing saw (Disco Automatic Dicing Saw 3220, Disco, Tokyo, Japan) was used to cut the entire wafer into small size SAW device chips. Then 6.2 μm Cu (copper) was electrochemically deposited on the exposed Au seed layer in an electrolyte environment. The second layer of the photoresist pattern was latter removed by acetone.

### Experimental measurement

A sinusoidal electric field with input voltage of 0.1 V and sweeping frequency from 35-95 MHz was applied to the IDT to excite a broadband input signal into the entrance port of the SAW waveguide using a signal generator (WF1967 multifunction generator, NF Corporation, Yokohama, Japan) and amplifier (ZHL-1-2W-S+, Mini-Circuits, Brooklyn, NY, USA). The actual voltage, current, and power across the device were measured using a digital storage oscilloscope (InfiniiVision 2000 X-Series, Keysight Technologies, Santa Rosa, CA). The source IDT is of the aperture of $1.44a$ (overlapping width) and is $5.2a$ away from the entrance port of the waveguide. To eliminate reflections at the boundaries of the device, a SAW absorber (Dragon Skin 10 Medium, Smooth-On, Inc., Macungie, PA, USA) is placed around the edge of the sample. The backside of the LiNbO$_3$ wafer is roughened to absorb possible reflection of the bulk acoustic wave at the bottom of the wafer. The out-of-plane displacement magnitude and phase fields are captured by laser Doppler vibrometer (LDV, UHF–120, Polytec, Waldbronn, Germany). The data presented is the average after 10 measurements from the LDV.

### Numerical Simulation

The eigen-mode and the driven-mode simulations were implemented using the commercial software COMSOL Multiphysics with the Acoustic (Acoustic-Solid Interaction) and Electrostatics modules, based on the finite element method. Floquet periodic boundaries were assigned for unit cell and supercell band diagram calculations, while the low-reflection boundary was imposed on the outer boundaries for the frequency domain driven-mode studies. A fixed boundary is always applied at the bottom of the LiNbO$_3$ substrate. On the band diagrams for the unit cells, the SAW modes can be distinguished under the sound cone, which is the formed by the slowest bulk mode dispersion. In the driven mode simulation, we excite the SAW by applying a sinusoidal edge load or a point load on the substrate. For the material properties, we used the z-cut LiNbO$_3$ parameters with a rotated coordinate system to get the properties for the 128-degree y/x-cut LiNbO$_3$ wafers. The elastic parameters of the Cu pillars used in the calculations are density $\rho_{Cu}$ = 8960 kg m$^{-3}$, Young's modulus $E_{Ni}$ = 70 GPa and Poisson's ratio $\nu_{Cu}$=0.34. Note that the Young's modulus is smaller than the conventional Young's modulus for Cu, due to our specific plating process. It was also found in the literature that Young's modulus can be sensitive to plating conditions[74].

For the Berry curvature calculation[75], the complex out-of-plane displacement with magnitude and phase information in the real-space domain is exported from COMSOL simulations for each wavevector for the integration.

## Data availability

The datasets within the article in the current study are available from the authors upon request.

## Acknowledgements

The authors thank Nirjhar Sarkar for discussion about the experimental settings. This work was supported by Army Research Office contract W911NF-17-1-0453.